# ARTICLE

# Phase-Transition-Driven Hyperbolic Optical Response and Directional Polaritons in Epitaxial VO$_2$ Thin Films

Maria Chiara Paolozzi,[a] Annalisa D'Arco,[a] Ilaria Martinelli,[a] Lorenzo Mosesso,[a] Jacopo Sera,[a] Alessandro D'Elia,[b] Augusto Marcelli,[c,d,e] Yingxue Chen,[f] Chongwen Zou,[f] Maria Cristina Larciprete,[g] Marco Centini,[g] Stefano Lupi [a,h], and Salvatore Macis*[a,h]

[a] Department of Physics, Sapienza University, Piazzale Aldo Moro 5, Rome, 00185, Italy.
[b] INFN - Laboratori Nazionali di Frascati, via Enrico Fermi 54, 00044, Frascati (Rome), Italy.
[c] Universita' di Camerino, Via Madonna delle Carceri 9, Camerino (MC) Italy
[d] INFN, Turin unit, Via Giuria 1, Turin 10125, Italy
[e] Rome International Centre for Material Science Superstripes, Via dei Sabelli 119A, 00185 Rome, Italy
[f] National Synchrotron Radiation Laboratory, School of Nuclear Science and Technology, University of Science and Technology of China, Hefei, 230029, PR China
[g] Dipartimento di Scienze di Base e Applicate per l'Ingegneria, Sapienza Università di Roma, Via A. Scarpa 16, Roma, 00161 Italy
[h] INFN - Sezione di Roma, Piazzale Aldo Moro, 2 - 00185 Rome, Italy

Optical anisotropy in crystalline solids enables direction-dependent light-matter interactions and underpins a variety of advanced photonic functionalities. In this context, Vanadium dioxide (VO$_2$) represents a prototypical material that undergoes a reversible Metal-to-Insulator Transition (MIT) near 67 °C, accompanied by pronounced electronic, structural, and optical modifications. The MIT not only dramatically modifies the VO$_2$ electrical conductivity but also reshapes its anisotropic optical response, making VO$_2$ an exceptional platform for dynamically tunable photonic and optoelectronic devices. In this work, we investigate how the intrinsic crystalline anisotropy of VO$_2$ induces a hyperbolic optical behavior in the metallic rutile phase. We study two epitaxial VO$_2$ thin films of different thicknesses grown on (110)-oriented MgF2 substrates. Broadband polarized spectroscopic measurements, extending from the infrared to the ultraviolet spectral range, are employed to independently investigate the optical response in both the monoclinic and rutile phases. From these measurements, we extract the optical conductivity and the dielectric function, revealing a pronounced anisotropy in the rutile metallic phase, with an enhanced free-carrier response along the rutile $c_R$ axis. Our data show that, within a narrow near-infrared spectral window, the real parts of the dielectric tensor components along the two principal axes acquire opposite signs, indicating the emergence of a hyperbolic type-II dispersion. The hyperbolic response is quantitatively evaluated through the quality factor and the degree of dielectric anisotropy, enabling a systematic assessment of VO$_2$ as a thermally switchable, hyperbolic optical medium. These findings expand the understanding of anisotropy-driven optical phenomena in phase-change materials and highlight VO$_2$ thin films as a promising platform for tunable and reconfigurable photonic applications.

## Introduction

Transition-metal oxides host a wide variety of emergent electronic phases arising from the interplay between strong electron correlations, lattice degrees of freedom, and orbital physics.[1-4] Within this class, vanadium oxides have long served as prototypical systems for investigating Metal-Insulator transitions (MITs) and exotic electrodynamics phenomena.[5] In particular, V$_2$O$_3$ has been extensively studied as an archetypal Mott system, where optical spectroscopy, angle-resolved photoemission, pressure- and field-dependent experiments have provided deep insight into orbital selectivity,[4-7] quasiparticle coherence, and the microscopic mechanisms driving the transition across its whole phase diagram.[7] Vanadium oxides have also enabled the exploration of non-equilibrium and field-driven phenomena, including terahertz (THz) electric-field-induced Mott transitions and pressure-driven metallization.[10,11] These studies have highlighted the sensitivity of correlated electronic phases to external perturbations and the central role of optical spectroscopy in tracking the evolution of low-energy electrodynamics. Within the vanadium-oxide broad family, $VO_2$ occupies a unique position since its discovery because it exhibits a reversible insulator-metal transition, accompanied by pronounced changes in its electrical, optical, and thermal properties, including a resistivity drop of approximately five orders of magnitude.[12] During the MIT, $VO_2$ undergoes a structural phase transition from a low-temperature monoclinic M1 (insulating) phase to a high-temperature tetragonal rutile R (metallic) phase, when heated above the critical temperature $T_c$=67 °C (about 340 K).[13] $VO_2$ continues to attract considerable interest because its MIT can be triggered also by multiple external stimuli, including pressure,[11] and applied voltage,[14] thereby enabling both fundamental studies and practical applications. The large and reversible property modulation associated with the MIT makes $VO_2$ a promising material for a wide range of technologies, from conventional optical switching and sensing devices to more advanced developments such as energy-efficient smart-window coatings, electronic components that exploit phase-change behaviour, ultrafast optical switches, neuromorphic memory elements, and adaptive thermal radiators.[15-18]

Understanding the mechanism governing the MIT, and thus the ability to tune its associated properties, is of relevant importance. Two primary interpretations have been proposed to explain how the transition occurs: the Mott-Hubbard mechanism, in which strong electron-electron interactions drive the phase transition, and the Peierls mechanism, where structural distortions modify the band structure and thereby induce the transition in $VO_2$.[19,20] Despite extensive research, the detailed microscopic origin of the MIT in $VO_2$ remains partially unresolved, and the respective roles of different degrees of freedom, such as orbital occupancy, lattice distortions, and electron correlations, have not yet been fully disentangled.[9-11] Nevertheless, it is well established that all these parameters can be used as effective tools to control the MIT. Among the various approaches, epitaxial strain and chemical substitution with higher or lower valence elements are the most widely employed methods to manipulate and tailor the MIT characteristics, including the transition temperature and hysteresis width.[11-13] For instance, chemical doping with tungsten[26-28] or molybdenum[29,30] is known to lower the transition temperature to room temperature or below, whereas chemical doping with titanium affects the electronic contribution to the optical conductivity in the rutile phase.[31]

Another important parameter to consider is the intrinsic electronic anisotropy of $VO_2$. Indeed, across the MIT, the band structure undergoes an abrupt reorganization: the band gap decreases sharply, dropping from approximately 0.67 eV in the monoclinic phase to 0 eV in the tetragonal phase. At the same time, the Fermi level crosses the $d_{//}$ and antibonding $\pi^*$ bands.[2,9,11,12,16-19] These bands are oriented along the $c$ axis and in the $ab$ plane of the tetragonal metallic phase, respectively, and their relative electron occupancy is a key factor governing the MIT. Despite this intrinsic electronic anisotropy, most studies of $VO_2$ transport and optical properties have focused on the $ab$ plane under various external perturbations, such as temperature variations[36], applied electric and magnetic fields[37], and sub-ps optical excitation[38], therefore a deep investigation of these properties along the $c$ axis is still lacking.

In this paper, we investigate the optical and electrical transport properties of thin $VO_2$ films epitaxially grown on $MgF_2$ substrates along the (110) direction. This orientation allows the growth of $VO_2$ films with both the $a_R$ and $c_R$ axes lying in the plane of the rutile phase, providing an ideal platform to systematically study the material's anisotropic behavior and its evolution with film thickness.

In particular, transmittance measurements were performed on films of varying thickness across a broad spectral range, from the infrared (IR) to the ultraviolet (UV), enabling a direct evaluation of the dielectric tensor $\epsilon$, which provides a clear evidence of an optical anisotropy between the $a_R$ and the $c_R$ axis. Our results further reveal an optical hyperbolic response of $VO_2$, with a near-infrared (NIR) hyperbolic operational spectral bandwidth (HOSB) in which the components of the dielectric tensor have opposite signs. Overall, this study not only provides a solid experimental evidence of the hyperbolic optical properties of $VO_2$ but also extends the catalogue of natural hyperbolic materials.

## Experimental Section

### Synthesis

Two $VO_2$ thin films were grown epitaxially on 0.5 mm x 5 mm x 10 mm crystalline (110) oriented $MgF_2$ substrates by Pulsed Laser Deposition (PLD) at the University of Science and Technology of China (USTC) in Hefei, China.[39-44] This substrate orientation allows growing the film on an unconventional crystal orientation, with the long axis of the rutile (tetragonal) phase in-plane. The deposition procedure was performed by irradiating a $VO_2$ ceramic target with a KrF pulsed laser at 248 nm and with a 10 Hz pulse frequency. The $VO_2$ ceramic target was prepared by pressing and sintering high-purity $VO_2$ powder (99.99% purity) in an inert gas atmosphere for more than twenty hours, with a base pressure of $4 \cdot 10^{-4}$ Pa. During the $VO_2$ film deposition, the substrate temperature was kept at 600 °C, and the oxygen pressure was controlled by a mass flow controller. During the deposition process, the target and the substrate were both rotated to improve the uniformity of the prepared films. The nominal thickness δ of the two films, estimated by the PLD deposition rate (3-4 nm/min) and later confirmed by optical analysis, is 14 nm and 55 nm.

### Characterization

Transmittance measurements were performed in the infrared (IR)-ultraviolet (UV) range as a function of temperature and light polarization.
Measurements in the IR range were carried out with a Bruker Vertex 70v spectrometer coupled with a Hyperion microscope, scanning from 1000 cm$^{-1}$ up to 8000 cm$^{-1}$ with a spectral resolution of 4 cm$^{-1}$. These measurements have been further extended to the VIS-UV spectral range through a JASCO V770 spectrometer (3125 cm$^{-1}$ up to 50000 cm$^{-1}$). A set of IR and VIS/UV polarizers has been used to control the electric field direction with respect to the sample axes.
As showed in Fig. S1, each sample has a short horizontal, and a long vertical side. These sides are respectively the $a$ axis and the $c$ axis of the $MgF_2$ substrate, and match the direction of the $a$ axis and $c$ axis of the $VO_2$ rutile phase. So in this work, we will refer to the polarization directions as parallel to the axis $a_R$ (horizontal direction) and $c_R$ (vertical direction).
Each measurement was performed at two temperatures, at 25 °C in the monoclinic phase, and above the $T_c$, at 110°C in the rutile phase.

The transmittance spectra of the two-layer systems composed of $VO_2$ film plus $MgF_2$ substrate were fitted with the RefFit software[45] (see Fig. S2 in Supplementary Information), which enables the modelling of the measured macroscopic optical quantities via a set of Drude-Lorentz oscillators, decoupling substrate and film optical responses. From this procedure, the $VO_2$ film optical conductivity and dielectric function were extracted along the $a_R$ and $c_R$ axis.

## Results and discussion

The four panels in Fig. 1 show the real part $\sigma_1(\omega)$ of the extracted optical conductivity in the IR-UV spectral range, extending from 1000 $cm^{-1}$ to 50000 $cm^{-1}$. More in detail, two different $VO_2$ film of 14 nm (panels a-c) and 55 nm (panels b-d) are displayed, and measured with two different polarizations, parallel to the $a_R$ (top) and $c_R$ axis (bottom), respectively. Each panel shows both monoclinic and rutile phases, in which several modulations associated with electronic transitions are detectable.

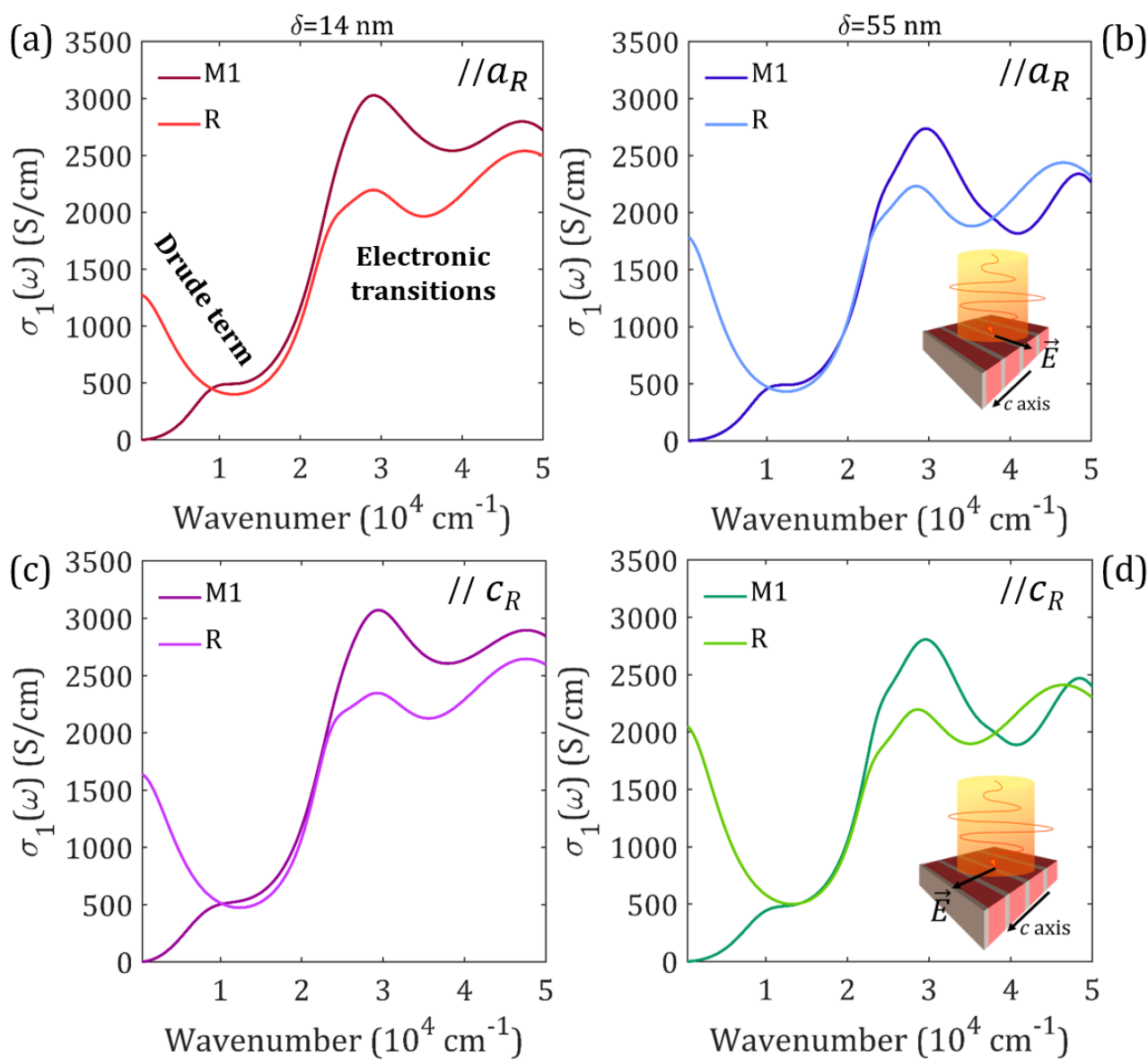

*Fig. 1* Real part of the optical conductivity $\sigma_1(\omega)$ in the IR-UV spectral range, extending from 1000 $cm^{-1}$ to 50000 $cm^{-1}$. Panels (a) and (c) report $\sigma_1(\omega)$ for the M1 and R phases (bordeaux and red curves) obtained with light polarization parallel to the $a_R$ axis, as highlighted in the sketch in panel (b), and for M1 and R phases (violet and fuchsia curves) obtained with light polarization parallel to the $c_R$ axis, as highlighted in the sketch in panel (d), for the 14 nm thick film. On the other hand, panels (b) and (d) report $\sigma_1(\omega)$ for the M1 and R phases (blue and cyan curves) parallel to the $a_R$ axis, and M1 and R phases (dark green and light green) parallel to the $c_R$ axis for the 55 nm thick film. Drude term and electronic transition contributions are highlighted in the first panel.

The first notable spectral feature in the M1 phase is the relatively weak peak observed at approximately 10000 $cm^{-1}$. This peak arises from optical transitions between the filled lower $d_{//}$ band and the empty $\pi^*$, and $d_{//}^*$ band across the optical gap.[46,47] The peak at approximately 23000 $cm^{-1}$ is related to transitions between the lower filled $d_{//}$ band and the upper empty band[48], and is stronger in the thicker sample (55 nm). The narrow nature of the $d_{//}$ bands, ascribed to the quasi-one-dimensional character of the V-V chains, helps explain why this feature is sharper when the electric field is parallel to the $a_R$ axis of the monoclinic phase ($c_R$ axis in the rutile phase).[46,49] The peak at ~ 29000 cm$^{-1}$ arises from transitions between the filled σ and π bands and the empty π* bands. The feature at approximately 38000 cm$^{-1}$, mainly visible in the thicker film, is related to transitions from the filled σ and π bands to the $d_{//}^*$ band. At last, the prominent peak at 48000 cm$^{-1}$ is determined by transitions between the occupied π orbitals and the empty σ* states.[48] Overall, the measured peaks are in good agreement with previous experimental data[46,48] and theoretical predictions.[49]

In the R phase, instead, a free carrier contribution is detectable up to 10000 cm$^{-1}$ with a characteristic Drude-type peak shape. Above the local minimum at 15000 cm$^{-1}$, features due to interband transitions appear. At ~ 24000 and 28000 cm$^{-1}$, the same transitions we described in the M1 phase from the filled σ and π bands to π* and $d^*_{//}$ bands are detected. In the metallic phase, the π* and $d_{//}^*$ bands are partially filled; therefore, the optical transitions have a reduced weight compared to their intensity in the monoclinic M1 phase, as can be noticed in each panel. Additionally, the peak at 48000 cm$^{-1}$ is shifted with respect to the monoclinic phase due to the rearrangement of π bands across the MIT.

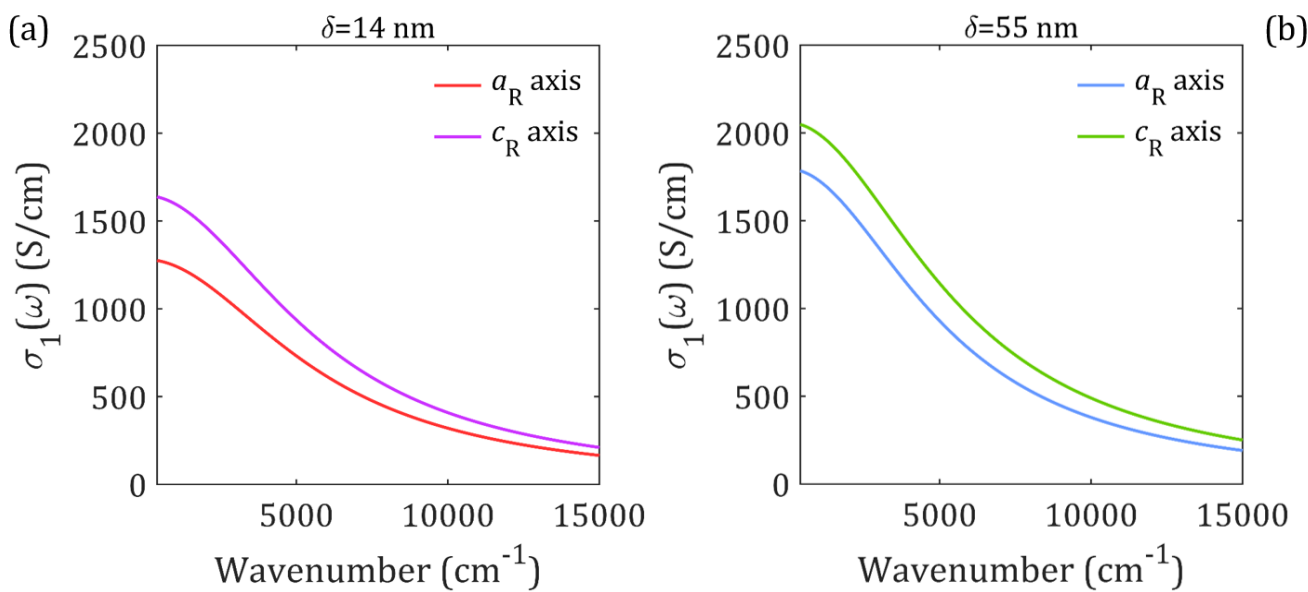

Fig. 2 Optical conductivity $\sigma_1(\omega)$ of the metallic phase (R) between 1000 cm$^{-1}$ and 15000 cm$^{-1}$ for the 14 nm (a) and 55 nm (b) thick VO$_2$ films. Electronic transition contributions were removed from the fit to emphasize the Drude term. Panel (a) reports the quantity for the 14 nm film along the $a_R$ (red curve) and $c_R$ (fuchsia curve) axes, whereas panel (b) shows the quantity for the 55 nm film along the $a_R$ (cyan curve) and $c_R$ (light green curve) axes.

Fig. 2 shows $\sigma_1(\omega)$ between 1000 cm$^{-1}$ and 15000 cm$^{-1}$ for the metallic R phase reported in Fig. 1. In this case, the electronic contributions associated with interband transitions were removed from the fit to emphasize the Drude-like terms in the 14 nm (a) and 55 nm cases (b), respectively. For both thicknesses, the Drude contribution is more intense along the $c_R$ axis, and overall is higher in the 55 nm thick film. This is evident also from the DC conductivity value $\sigma_0$ extracted from the fitting procedure at ω=0, and from the spectral weight obtained from the Drude plasma frequency $\omega_P$ (SW=$\omega_P^2$/8) along the $a_R$ and $c_R$ presented in Table 1. As already pointed out, the quantity is larger along the $c_R$ axis, and for a given polarization, it increases with film thickness. Interestingly, the SW ratio, defined as SW$_c$/SW$_a$, approaches a constant value of approximately 1.3 for each specific thickness.

Table 1. DC conductivity value $\sigma_0$ extracted from the fitting procedure and calculated spectral weight SW along the $a_R$ and $c_R$ axis for 14 and 55 nm thick VO$_2$ films in the rutile phase.

These two different Drude responses along the two polarization axes indicate a potential hyperbolic dispersion of these films. In the metallic R phase, along both $a_R$ and $c_R$ axes, the real part of the dielectric function $\epsilon_1$ is negative up to the NIR region. However, along the $a_R$ axis, where the Drude contribution is weaker, $\epsilon_{1,a_R}$ becomes positive at a certain frequency, while $\epsilon_{1,c_R}$ is still negative, satisfying the hyperbolicity condition of $\epsilon_{1,a_R} \cdot \epsilon_{1,c_R} < 0$. Given its crystalline structure, the VO$_2$ optical response in the Rutile phase along the crystallographic axes $a_R$ and $b_R$ is the same. This means that, in the hyperbolic region, VO$_2$ has two elements of the dielectric tensor with a positive real part ($\epsilon_{1,a_R}$ and $\epsilon_{1,b_R}$) and one with a negative real part ($\epsilon_{1,c_R}$), making it a type-I hyperbolic material.

|  | $\sigma_{0,\,aR}$ (S/cm) | $\sigma_{0,\,cR}$ (S/cm) | SW$_a$ (cm$^{-2}$) | SW$_c$ (cm$^{-2}$) |
|---|---|---|---|---|
| 14 nm | 1270 | 1600 | 5.5 · 10$^7$ | 7.2 · 10$^7$ |
| 55 nm | 1800 | 2000 | 7 · 10$^7$ | 8.6 · 10$^7$ |

Fig. 3a shows $\epsilon_1(\omega)$ along $a_R$ and $c_R$ directions in the R phase for δ=14 nm, restricted to the 7500 cm$^{-1}$ - 9000 cm$^{-1}$ spectral range to emphasize the region where hyperbolicity occurs. As can be noticed, $\epsilon_1(\omega)$ display negative values up to 7680 cm$^{-1}$ (8900 cm$^{-1}$) along $a_R$ ($c_R$) direction, where they cross the $\epsilon_1(\omega)=0$ line, and become positive. The light-orange shaded band marks the spectral window in which $\epsilon_{1,a_R} \cdot \epsilon_{1,c_R} < 0$, spanning the NIR region, with an operational spectral bandwidth of Δω=1220 cm$^{-1}$. In this spectral range, the hyperbolic response is governed by free carriers, that is, by a Drude contribution dominating along both axes. At higher frequencies, Lorentz oscillators associated with electronic transitions produce an increase in $\epsilon_1(\omega)$, which becomes and remains positive up to 50000 cm$^{-1}$ for both polarizations (see Fig. S3 in Supplementary Information).

Fig. 3b shows $\epsilon_1(\omega)$ along $a_R$ and $c_R$ directions in the R phase for δ=55 nm, once again restricted to the 7500 cm$^{-1}$ - 9000 cm$^{-1}$ spectral range. A similar behaviour is observed also for this thicker film, except for the hyperbolic region (marked by the light-pink shaded band), spanning the NIR between 8425 cm$^{-1}$ and 8840 cm$^{-1}$, with a HOSB of Δω=415 cm$^{-1}$.

To evaluate the hyperbolic performance of VO$_2$, two key parameters can be calculated. The first is the quality factor Q=-$\epsilon_1(\omega)/\epsilon_2(\omega)$, defined as the ratio between the real part $\epsilon_1$ and the imaginary part $\epsilon_2$ of the dielectric function along the most conductive axis ($\epsilon_1(\omega) <$ 0), and the second is the strength of the dielectric anisotopy (SDA), defined as the difference $\Delta\epsilon_1 = \epsilon_{1,a_R} - \epsilon_{1,c_R}$ between the real part of the dielectric function along the conductive $a_R$ axis ($\epsilon_{1,a_R}$) and that in the conductive $c_R$ axis ($\epsilon_{1,c_R}$).

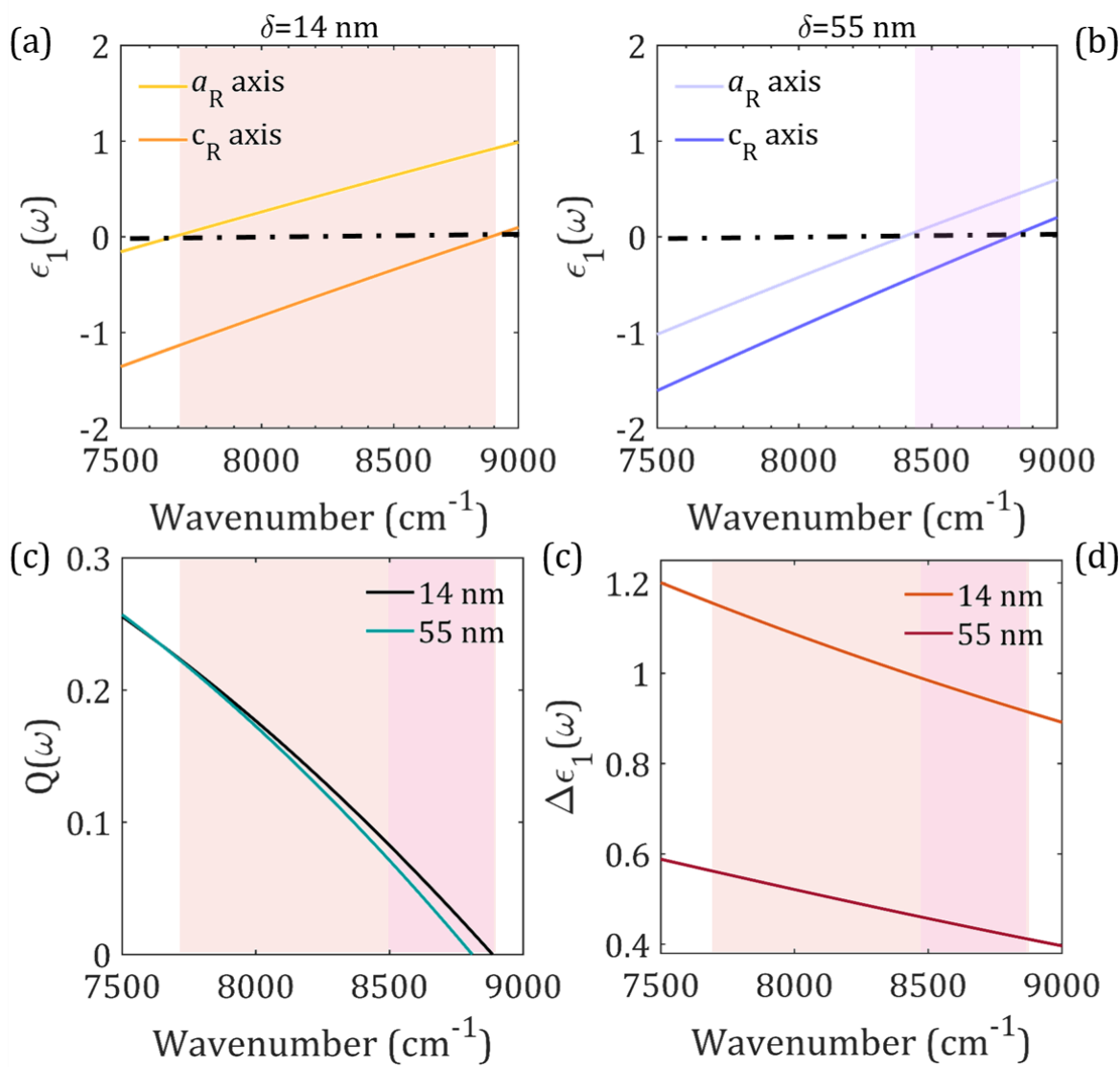

*Fig. 3 Quality factors describing the hyperbolic properties of the R phase of the VO$_2$ films with δ=14nm and 55 nm. (a) Real part of the dielectric function $\epsilon_1(\omega)$ along $a_R$ (yellow curve) and $c_R$ (orange curve) directions, restricted to the 7500 cm$^{-1}$ - 9000 cm$^{-1}$ spectral range, for δ=14 nm. The dashed black line shows where $\epsilon_1(\omega)=0$. (b) Real part of the dielectric function $\epsilon_1(\omega)$ along $a_R$ (lilac curve) and $c_R$ (purple curve) directions, restricted to the 7500 cm$^{-1}$ - 9000 cm$^{-1}$ spectral range, for δ=55 nm. (c) Quality factor Q(ω) evaluated between 7500 cm$^{-1}$ and 9000 cm$^{-1}$ for 14 nm (black curve) and 55 nm (teal green curve) thick films. (d) Strength of the dielectric anisotropy Δ$\epsilon_1(\omega)$ evaluated between 7500 cm$^{-1}$ and 9000 cm$^{-1}$ for 14 nm (dark orange curve) and 55 nm (bordeaux curve) thick films. The shaded, light-orange and light-pink bands in all the panels mark the spectral range where the hyperbolic condition is satisfied for the two films.*

Both parameters are reported in Fig. 3c-d for the 14 nm and 55 nm films between 7500 cm$^{-1}$ and 9000 cm$^{-1}$. The Q factor (Fig. 3c) decreases across the operational spectral bandwidth Δω, ranging from 0.23 to nearly 0 for the 14 nm thick film (black curve inside the light-orange band), and from 0.09 to 0 for the 55 nm thick film (teal green curve inside the light-pink band).

A similar trend is observed for the SDA (Fig. 3d), which decreases from about 1.16 to 0.91 for the 14 nm thick film (dark orange curve inside the light-orange band), and from 0.47 to 0.42 for the 55 nm thick film (bordeaux curve inside the light-pink band) within Δω.

The previously discussed experimental optical characteristics of VO$_2$ thin films also give rise to distinctive surface polaritonic behavior in the hyperbolic regime, as also partially discussed in similar systems.[50,51] In the presence of a strongly anisotropic dielectric response, hyperbolic materials are known to support surface plasmon polaritons with unconventional dispersion relations, commonly referred to as hyperbolic surface plasmon polaritons (HSPPs). These excitations combine the localized nature of conventional surface plasmon polaritons with the highly directional propagation imposed by hyperbolic dispersion, enabling access to large in-plane wavevectors and subwavelength electromagnetic confinement.

To visualize the hyperbolic nature of these polaritonic modes, we numerically calculated the in-plane isofrequency map of the 14nm thick VO$_2$ film in the metallic R phase, at a frequency of 8000 cm$^{-1}$. The map was obtained calculating the imaginary part of the p-polarized Fresnel reflection coefficient, Im(r$_{pp}$), evaluated for evanescent waves with in-plane momentum exceeding the light line. Calculations were performed within a transfer-matrix formalism for anisotropic, dissipative media, assuming a semi-infinite VO$_2$ layer with its $a_R$ and $c_R$ crystallographic axes lying in the plane of the interface with an air layer as sketched in Fig. 4a.[52-56] Fig. 4b displays the modulus of Im(r$_{pp}$) plotted in reciprocal space (k$_x$,k$_y$), normalized to the free-space wavevector $k_0$. This map represents cross-sections of the hyperbolic isofrequency surface and reveals open, strongly anisotropic contours characteristic of hyperbolic dispersion. The central circular region corresponds to the isotropic dispersion in air, inside which no surface modes can propagate. The form of the hyperbola is the one typical of type-I hyperbolic material. In analogy with other hyperbolic systems, the polaritonic response of VO$_2$ depends on the orientation of the incidence plane relative to the crystal axes. We define φ (see Figure 4a) as the angle between the plane of incidence and the $a_R$ axis of the material: to excite surface polaritons at a certain frequency, incident radiation has to match this angle φ.

From Fig.4b we can see that the hyperbola's asymptotes form with its center an angle of 30°, that is exactly the value of the angle φ needed to observe, on the VO$_2$ surface, polaritonic modes at a frequency of 8000 cm$^{-1}$.

This map helps to understand the concept of directionality: given that, in general, electric field group velocity $v_g$ is orthogonal to the isofrequency curve, surface polaritons in a hyperbolic medium, respect to a conventional one, propagate in a focused direction that depends on the form of the hyperbola, and so, on the frequency.

The ability to thermally and electrically activate and deactivate hyperbolic surface plasmon polaritons across the metal-insulator transition represents a unique functionality of $VO_2$, enabling reversible control of highly directional surface-wave propagation without the need for nanostructuring or external gating.

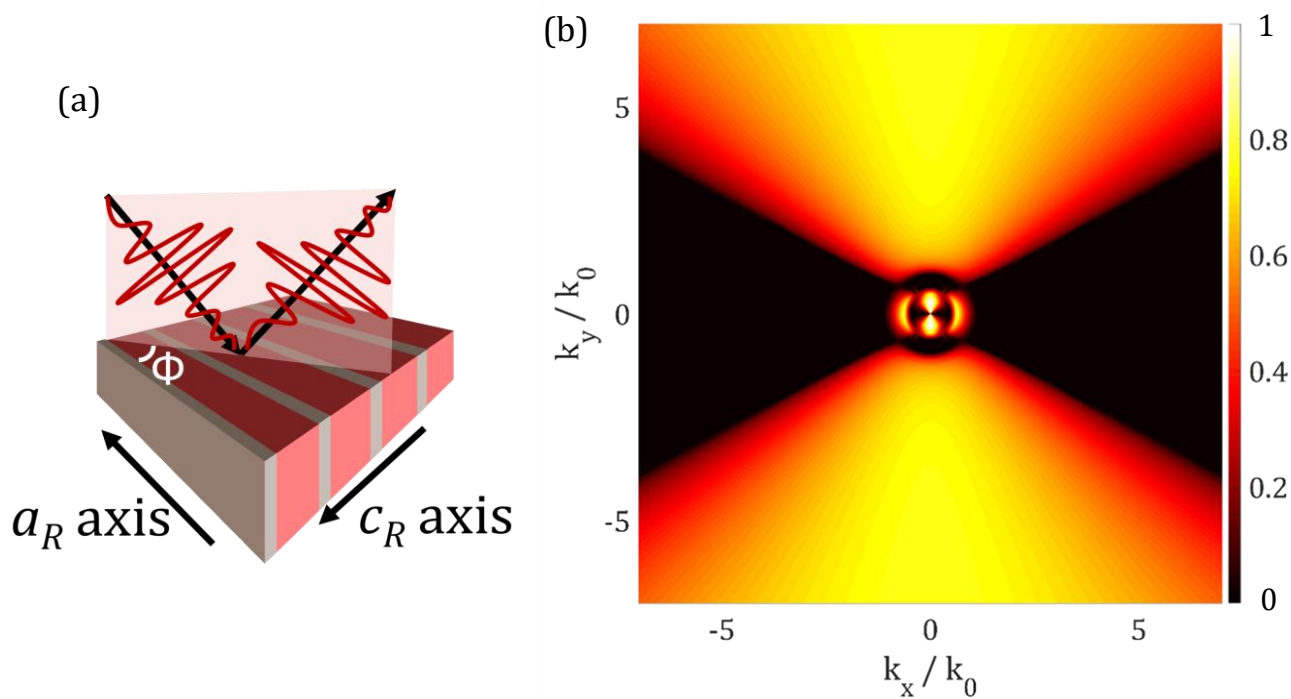

*Fig. 4 Panel (a) shows a schematic representation of the radiation impinging on the VO$_2$ surface, highlighting the angle φ formed by the radiation plane of incidence with the $a_R$ axis. Panel (b) reports the isofrequency curve of VO$_2$, calculated for an incident radiation frequency of 8000 cm$^{-1}$ as a function of the x and y components of the wave vector normalized with the modulus of the incident radiation $k_0$.*

Moreover, hyperbolic dispersion enables the propagation of modes with high wavevector magnitudes, and therefore very short wavelengths. For this reason, surface polaritonic modes can be confined to extremely small spatial scales, paving the way for various applications, such as hyperlenses capable of overcoming the diffraction limit, or optical circuits.

## Conclusions

In conclusion, we have investigated the anisotropic optical response of epitaxial $VO_2$ thin films grown on (110)-oriented $MgF_2$ substrates across the insulator-to-metal transition. Polarization-resolved spectroscopy over a broad spectral range, from the Infrared to the Ultraviolet, enabled the independent determination of the optical conductivity and dielectric function along the $a_R$ and $c_R$ crystallographic axes in both the monoclinic insulating and rutile metallic phases. In the metallic state, a pronounced anisotropy of the free-carrier response was observed, with higher optical and DC conductivity along the $c_R$ axis, reflecting the intrinsic directional nature of the $VO_2$ electronic structure. Beyond conventional anisotropy, we demonstrated that $VO_2$ exhibits Type-I hyperbolic optical behaviour within a spectral window in the NIR region, in the rutile phase, where the real parts of the dielectric tensor components along the two in-plane axes acquire opposite signs. The hyperbolic response is governed by free carriers and was quantitatively evaluated using the quality factor and the strength of the dielectric anisotropy, finding a better performance for the 14 nm thick film compared to the 55 nm. Although the operational bandwidth and quality factors are smaller than those of high-performance natural hyperbolic materials such as delafossite oxides, $VO_2$ uniquely combines hyperbolicity with a reversible, thermally driven phase transition. These findings establish epitaxial $VO_2$ thin films on (110) $MgF_2$ as a natural, switchable hyperbolic medium and provide a comprehensive experimental quantification of their anisotropic and hyperbolic optical properties, which could be exploited to build the first reconfigurable hyperbolic devices. Further optimization through strain engineering, thickness control, or electrostatic tuning may significantly enhance its hyperbolic performance.

## Author contributions

Maria Chiara Paolozzi: writing - original draft, writing - review & editing, investigation, formal analysis. Annalisa D'Arco: writing - review & editing. Ilaria Martinelli: writing - review & editing, investigation. Lorenzo Mosesso: writing - review & editing, investigation. Jacopo Sera: writing - review & editing, simulations. Alessandro D'Elia: writing - original draft, writing - review & editing. Augusto Marcelli: writing - review & editing. Ze Yu Wu: writing - review & editing, materials. Stefano Lupi: writing - original draft, writing - review & editing, supervision, validation, methodology. Salvatore Macis: writing - original draft, writing - review & editing, conceptualization, supervision, validation, methodology.

## Conflicts of interest

In accordance with our policy on Conflicts of interest please ensure that a conflicts of interest statement is included in your manuscript here. Please note that this statement is required for all submitted manuscripts. If no conflicts exist, please state that "There are no conflicts to declare".

## Data availability

A data availability statement (DAS) is required to be submitted alongside all articles. Please read our full guidance on data availability statements for more details and examples of suitable statements you can use.

## Acknowledgements

This publication was supported by the European Union under the Italian National Recovery and Resilience Plan (NRRP) of Next Generation EU partnership PE0000023-NQSTI.

## Notes and references